# Neutrino Telescope in Lake Baikal: Present and Nearest Future

*I. Belolaptikov[a] and Zh.-A. Dzhilkibaev[b],*** on behalf of the Baikal-GVD Collaboration
(a complete list of authors can be found at the end of the proceedings)

[a] *Joint Institute for Nuclear Research,*
   *Joliot Curie 6, Dubna, Russia*

[b] *Institute for Nuclear Research, Russian Academy of Sciences,*
   *60th Anniversary of October Avenue 7a, Moscow, Russia*

   *E-mail:* igor.belolaptikov@jinr.ru, djilkib@yandex.ru

The progress in the construction and operation of the Baikal Gigaton Volume Detector in Lake Baikal is reported. The detector is designed for search for high energy neutrinos whose sources are not yet reliably identified. It currently includes 2304 optical modules arranged on 64 strings, providing an effective volume of 0.4 km$^3$ for cascades with energy above 100 TeV. We review the scientific case for Baikal-GVD, the construction plan, and first results from the partially built experiment, which is currently the largest neutrino telescope in the Northern Hemisphere and still growing up.

---

*Presenter

1. Introduction

High-energy neutrino astronomy, a very recent and lively research field, has emerged by a construction of gigaton volume neutrino telescopes deep in ice and under water at both the Southern (IceCube) and Northern (Baikal-GVD, KM3NeT) Hemispheres.

The investigation of astrophysical high-energy neutrino flux gives additional information on production mechanisms of radiation in the Universe. Neutrinos, CR and high energy γ-rays can be produced at the sources or in the encountered dense source environment. Unlike CRs, which in their path towards the Earth are deflected by the intergalactic magnetic fields or absorbed as they interact with matter or radiation via inelastic collisions, neutrinos travel undisturbed. High energy γ-rays ($E > \sim 300$ TeV) are absorbed by cosmic microwave background radiation. Only by combing information from all three messengers, one will be able to reflect a complete picture of processes in the Universe. Gravitational Waves, which were recently detected, are one additional probe to investigate energetic phenomena in Universe.

IceCube has approved the existence of neutrinos of cosmic origin with the detection of a diffuse flux exceeding the expected background. The origin of these neutrinos is yet unknown. Only due to the simultaneous observations with other astrophysical messengers, the first neutrino source has been discovered [1, 2]. This fact points us to multi-messenger astronomy as most promising method for searching new sources of neutrino of astrophysical nature.

The Baikal Gigaton Volume Detector (Baikal-GVD) [3] is located in the south part of Lake Baikal, where the lake depth is 1366 m. The detector uses 10-inch photomultiplier tubes (PMTs) to detect the Cherenkov light from charged particles produced in neutrino interactions. The detector elements are arranged along vertical strings, which are in turn arranged in heptagonal clusters. Each cluster has its own connection to the shore station and acts as an independent detector. Eight clusters, with a total of 2304 PMTs, have already been deployed. Six more clusters are scheduled for deployment in the next three years. Baikal-GVD is aimed to provide observations of the TeV-PeV neutrino sky with a sensitivity similar to that of IceCube [4] and KM3NeT-ARCA [5]. The Baikal-GVD site offers very convenient location for the observation of the Southern sky, where the Galactic center and most part of the Galactic Plane can be investigated. The detector has opportunity to observe the sky "in neutrino" from a complimentary field of view of IceCube.

2. The Baikal-GVD neutrino telescope

The Baikal-GVD detector is installed 3.6 km offshore in the southern basin of Lake Baikal at 51° 46'N and 104° 24'E. The lake depth at the detector location is nearly constant at 1366-1367 m below the nominal surface level of the lake. The detector elements are arranged along vertical strings, each of which is anchored to the bottom of the lake and kept taut by a bunch of buoys at the top (see Fig. 2.1). Each string holds 36 optical modules. The optical module (OM) comprises a 10-inch high-quantum-efficiency PMT (Hamamatsu R7081-100), a high voltage unit and front-end electronics, all together enclosed in a pressure-resistant glass sphere. The OM is also equipped with calibration LEDs and various digital sensors, including an accelerometer/tiltmeter, a compass, a pressure sensor, a humidity sensor, and two temperature sensors. The OMs are arranged with 15 m vertical spacing, for a total active string length of 525 m, starting 90 m above the lakebed. The PMT photocathodes are oriented vertically downwards. The OMs are mechanically attached to a load-carrying cable. Additionally, each string has four electronics modules, also housed inside glass spheres. Three of these modules are Section Modules, each of which serves a group of 12 OMs. The Section Modules provide power to the OMs and digitize the PMT signals with a 200 MHz frequency. The OMs are connected to their respective Section Modules via electric cables that are laid along the load-carrying cable. The fourth electronics module is the String Module. The String Module acts as a hub for power distribution and communication with the Section Modules. The string also holds hydrophones for acoustic monitoring of the PMT positions [6] and LED beacons for detector calibration [7, 8].

The strings are grouped in clusters, with 8 strings per cluster, as shown in Fig. 2.1. The positions of the strings on the lakebed form a heptagon with one central string and seven peripheral strings, with an average horizontal spacing between the strings of ~ 60 m. The eight strings are connected to the Cluster Center assembly, which is comprised of four connected units, installed on a common installation plate attached to the load-carrying cable of the central string of the cluster at a shallow depth. The Cluster Center is connected to the shore station via a dedicated electro-optical cable. The Cluster Center is responsible for distributing the time synchronization signals to the individual strings, cluster-level triggering and data transmission to shore. The standard trigger condition requires two neighboring channels within the same section (of 12 OMs) to be hit within a 100 ns time window, with a minimal requirement for the hit amplitude $A_{high}$ for one of the hits and $A_{low}$ for the other. $A_{high}$ and $A_{low}$ slightly vary from channel to channel. Mean values are $A_{high} \approx 3.5$ photo-electrons (p.e.) and $A_{low} \approx 1.7$ p.e. Once this trigger condition is met for any of the Section Modules, a 5 μsec event time frame is read out from all the Section Modules of the cluster. Thus, each cluster can operate as a stand-alone neutrino detector. The effective volume of a GVD cluster for cascade-like neutrino events with energy above 100 TeV is estimated as 0.05 km$^3$ [9].

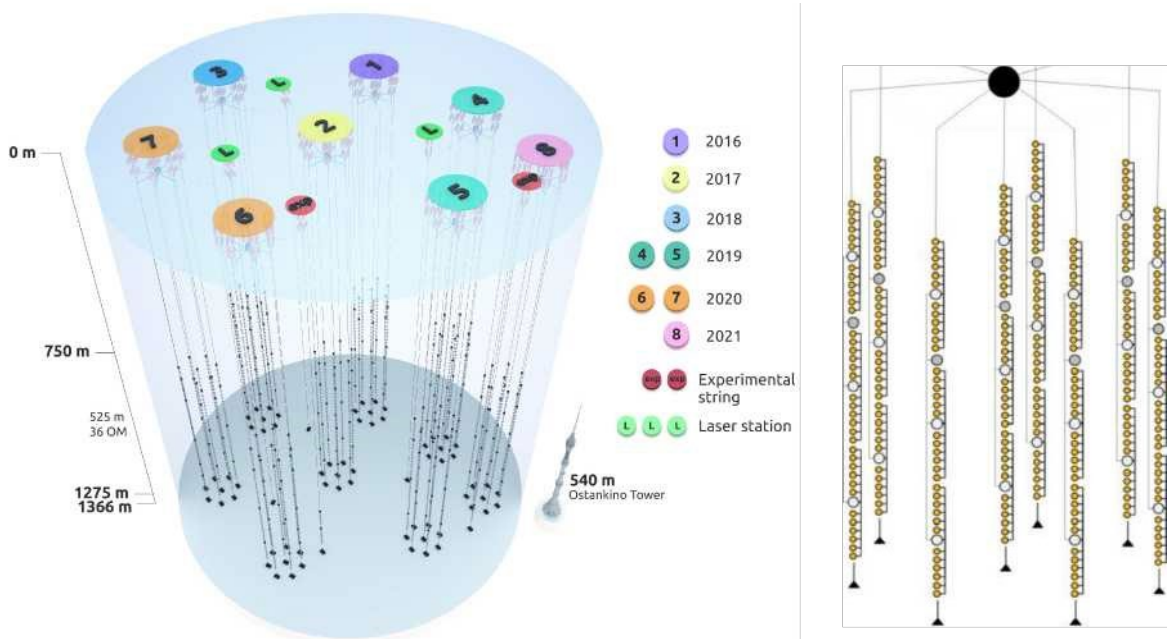

Figure 2.1: Left: Schematic view of the Baikal-GVD detector. The yearly progression of the detector deployment is shown in the legend. Right: The Baikal-GVD cluster layout (vertical scale compressed).

The clusters are arranged on the lakebed in a hexagonal pattern, with a distance ~ 300

meters between the cluster centers. A common synchronization clock allows for subsequent merging of the physics event data collected from the different clusters. Additional technical strings equipped with high-power pulsed lasers are installed in-between the GVD clusters. These are used for detector calibration [8] and light propagation studies [10]. The lake is covered with thick ice (up to ~ 1m) from February to mid-April, providing a convenient solid platform for detector deployment and maintenance operations.

According to a study made with a specialized device, the light absorption length in the deep lake water reaches maximal values, ~ 24 m, at a wavelength of 488 nm [11]. The effective light scattering length is ~ 480 m (at 475 nm; see [11] for details). Both the absorption and scattering characteristics show variations with depth and over time.

The optical modules detect the Cherenkov light from secondary charged particles resulting from neutrino interactions. The times of the pulses are used to reconstruct the neutrino direction, and the integrated charges (or amplitudes) provide a measure of the neutrino energy. The detector layout is optimized for the measurement of astrophysical neutrinos in the TeV-PeV energy range. Events resulting from charged current (CC) interactions of muon (anti-)neutrinos will have a track-like topology, while the CC interactions of the other neutrino flavors and neutral current (NC) interactions of all flavors will typically be observed as nearly point-like events. Hence, the observed neutrino events are classified into two event classes: tracks and cascades.

The first cluster of Baikal-GVD was deployed in 2016. Two more clusters were added in 2017 and 2018, followed by two more in 2019, another two in 2020, and one more in 2021. As of April 2021, the detector consists of 8 clusters, occupying a water volume of ~ 0.4 km$^3$. As it stands, Baikal-GVD is currently the largest neutrino telescope in the Northern Hemisphere. The construction plan for the period from 2022 to 2024 anticipates the deployment of six additional GVD clusters.

All Baikal-GVD clusters generally show stable operation. Occasional failures of individual optical or electronics modules, e.g. due to water leaks, are fixed during the regular winter campaigns. Each detector string can be recovered and re-deployed without the need to recover the whole cluster.

## 3. Data acquisition system

Work on the full-scale deployment of the telescope began in 2015. From that moment on, the configuration of the data acquisition system (DAQ) as a whole remained unchanged. However, the experience of operating the installation has shown the need for partial modernization of individual detector units. In particular, in order to increase the reliability and efficiency of detector operation, changes were made to the power supply, calibration and positioning systems, and deep-sea cable network was optimized.

### 3.1 Structure of the data acquisition system

The structure of the DAQ is determined by the configuration of the telescope and is formed from electronic control modules for clusters, strings and sections of optical modules [12]. The section is the base unit of the DAQ, which includes 12 OMs, 2 acoustic modems (AM) of the positioning system and a Section Module (see Fig. 3.1.1). Optical modules and acoustic modems are connected to a Section Module, whose functions are the control, collection and primary processing of OM data. The connection is carried out by individual cables with a length of about 90 meters. The power supply of OMs and AMs is provided by a 16-channel DC commutator: 12 channels are served to optical modules (12 V), 2 channels are used for acoustic modems (24 V), two channels are reserved for system expansion. The DC commutator unit allows also to control the operating modes of the OMs and monitor their parameters. The control is carried out via six switchable channels of the RS-485 bus: each channel serves two OMs. Ethernet COM Server NPort IA 5250 is used to control the operation of the AMs.

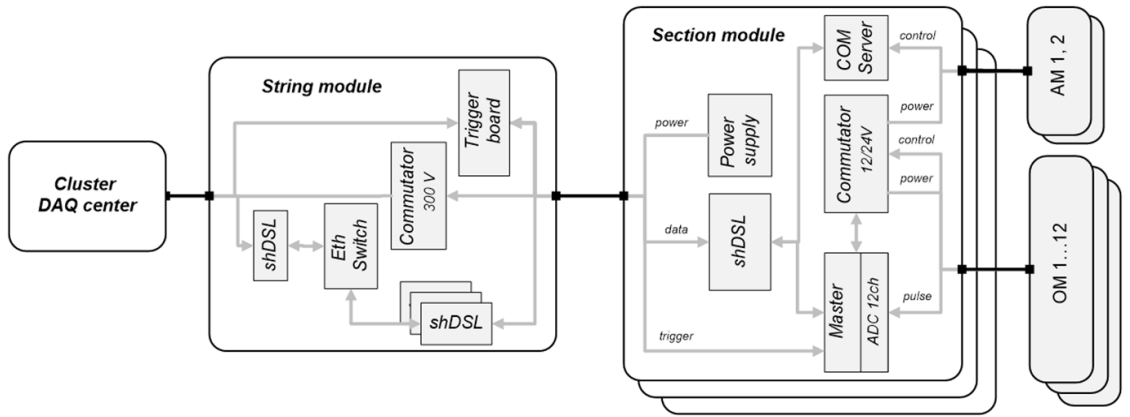

Figure 3.1.1. The block diagram of the cluster data acquisition system.

The processing of analog signals coming from the OMs is carried out by the Master unit. The unit includes a 12-channel ADC with a sampling rate of 200 MHz, which provides continuous scanning of OM outputs and recording information in a 30 microseconds cyclic buffer. Data processing (the formation of the section request and the event frame) is performed at the level of FPGA Xilinx Spartan 6.

The condition for forming a section request can be either coincidences of signals from two neighboring OMs (basic operation mode), or a majority coincidences of OMs signals (test mode of operation). A common trigger of the cluster initiates the formation of an event frame of 5 microseconds duration. The time frames of the sections are transmitted via the IEX-402-SHDSL Ethernet Extenders to the shore station (transmission speed is up to 5.7 Mbit/s). To increase the bandwidth of the channel, data is filtered. Only those parts of the time frame that contain signals exceeding the specified threshold are transmitted.

The String Module controls the operation of three sections. Each section is connected to the String Module via a separate cable. A 12-channel 300 VDC commutator is installed in the String Module to power the sections. Six channels are used for the power supply of 3 sections (the OMs and Master units of each section are powered through separate channels). Six channels are used as backup. The channel switching is controlled by digital output module ICP DAS I-7045. The analog input module ICP DAS I-7017Z-G is used to monitor the output voltage on the power channels. The request signals coming from the three sections are combined on the string trigger board to form a string request.

Power supply to eight strings is provided by two 300 VDC commutators of the Cluster Center: the main and backup. The trigger system of the Cluster Center is designed on the basis of the Master unit. The string requests are sent to the ADC inputs of the Master, which forms a common cluster trigger. Currently, a common trigger is generated for each incoming string request. At the same time, the mode of requests coincidences from two or more strings is foreseen. A common trigger synchronizes the operation of all sections of the cluster and initiates data reading of the sections. The data from the strings is transmitted to the Shore Center via a fiber-optic hybrid cable with a length of about 7 km (transmission speed of 1 Gbit/s).

### 3.2 Time calibration system

Baikal-GVD time calibration [7,8] consists in measuring the relative time delays of signals on the channels using calibration light sources. Light sources developed on the basis of Kingbright L7113 LEDs with a wavelength of 470 nm at the maximum radiation and a pulse duration of ~5 ns are used For time calibration. The intensity of their radiation is regulated from units of photons to $10^8$ photons per flash. The light pulse is formed in a cone with a width of 15° and can be registered by an optical modules at distances up to 100 m from the radiation source. Each optical module has two calibration sources, with LEDs oriented in the upper direction. In addition, in the 12 OMs of the cluster, placed on the central and two peripheral strings, two matrices of 5 LED sources are installed. LEDs are oriented horizontally and arranged evenly around the circle. Figure 3.2.1 shows examples of events initiated by flashes of the calibration LED and the LED matrix. The LEDs of the optical modules allow calibrating the time delays of the channels within a single string. LED matrices provide relative time calibration of channels of different strings. The accuracy of the time calibration is 2-3 ns.

Laser light sources are used for relative time calibration of the clusters. The lasers emit at a wavelength of 532 nm, the flash duration is about 1 ns, the maximum radiation intensity is about $10^{15}$ photons. A laser beam is introduced through a light guide into a diffuser forming point quasi-isotropic light source. The lasers are mounted on buireps, forming laser stations. In the basic configuration, laser station consists of two lasers. Laser stations are placed between clusters and provide mutual calibration of a group of 3 - 4 clusters. In addition to the calibration function, lasers are used to monitor the parameters of the water environment of Lake Baikal in the installation area.

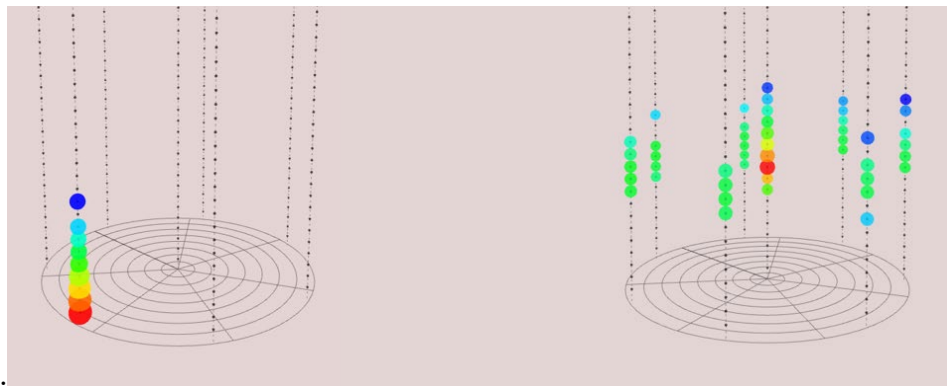

Figure 3.2.1. Examples of calibration events from the LED of the lower optical module of the section (left) and the LED matrix installed on the central garland (right).

### 3.3 Characteristics of the data acquisition system

The basic characteristics of the data acquisition system (the registration thresholds and accuracy in time measurement) are determined by the bandwidth of underwater network and parameters of the electronic units responsible for signal convertion and data processing. The bandwidth of the network is limited by the speed of the Ethernet Extenders shDSL (5.7 Mbit/s for each string). Each time frame formed by the sections contains information about the pulse shape from OMs. Stable transmission of this information to the shore station is possible at the cluster trigger rate not exceeding 200 Hz. This restricts the minimum trigger thresholds of the section on the level about 1.5 and 4 p.e. (see section 2). Such thresholds correspond to the frequency of trigger formation of 30 - 150 Hz, depending on the light activity of Lake Baikal.

The accuracy of measuring the time difference between triggered channels depends on the channels time accuracy and the accuracy of their synchronization. The Baikal-GVD synchronization system includes two subsystems that provide intra cluster channel synchronization, and sync between different clusters. Sync within a single cluster is achieved with a common trigger signal received by all sections in the cluster. To sync the clusters with each other, a single clock frequency is used, formed by the equipment at the shore station. The measurements are carried out by two independent systems: White Rabbit [13] and specially designed for the Baikal-GVD *SSBT*.

In order to study the sync accuracy, we analyzed the measurements of laser series which illuminated multiple clusters. The standard deviation (RMS) of the measured difference in the channel response time was used for accuracy estimation. For the channels within one Section, the measured RMS was 0.25 ns. For different sections, the RMS was $2.1 \pm 0.05$ ns, which is in a good agreement with the expected value of 2.04 ns. The spread of the formation time of cluster triggers with *WR* and *SSBT* is about 4 ns, which determines the accuracy of the absolute time binding of events. The intercluster synchronization system allows you to measure the timing of trigger formation with subnanosecond resolution, which ensures the accuracy of measuring the relative time between channel responses on different clusters of about 2 ns (RMS of time distribution), similar to intracluster intersection synchronization.

## 4.    Selected results

### 4.1 Observations of track-like neutrino events

Events resulting from charged current (CC) interactions of muon (anti-)neutrinos, as well as τ-neutrino CC interactions followed by leptonic τ decays, have a track-like topology in Baikal-GVD. A fast $\chi^2$-based reconstruction algorithm has been developed to reconstruct such track-like events [14]. The algorithm has been applied to a combined dataset of single-cluster events collected from the first five operational clusters of Baikal-GVD in April–June 2019 with a total single-cluster equivalent livetime of 323 days. This resulted in 9.8 million reconstructed events, most of which are due to atmospheric muons. A simple cut-based analysis has been employed in order to select neutrino events coming from below horizon and suppress the background of mis-reconstructed atmospheric muon events. The event selection cuts have been optimized for the detection of atmospheric neutrinos. The analysis yielded 44 neutrino candidate events, in excellent agreement with expectations. The observed zenith angle and energy distributions of the neutrino events also are in good agreement with Monte Carlo predictions (D.Zaborov, this conference, PoS(ICRC2021)1177). A mutli-cluster analysis using the same reconstruction algorithm has also been prepared, currently pending a final decision on the multi-cluster calibration readiness. Further analysis developments promise to improve the neutrino detection efficiency, angular resolution, and energy measurements.

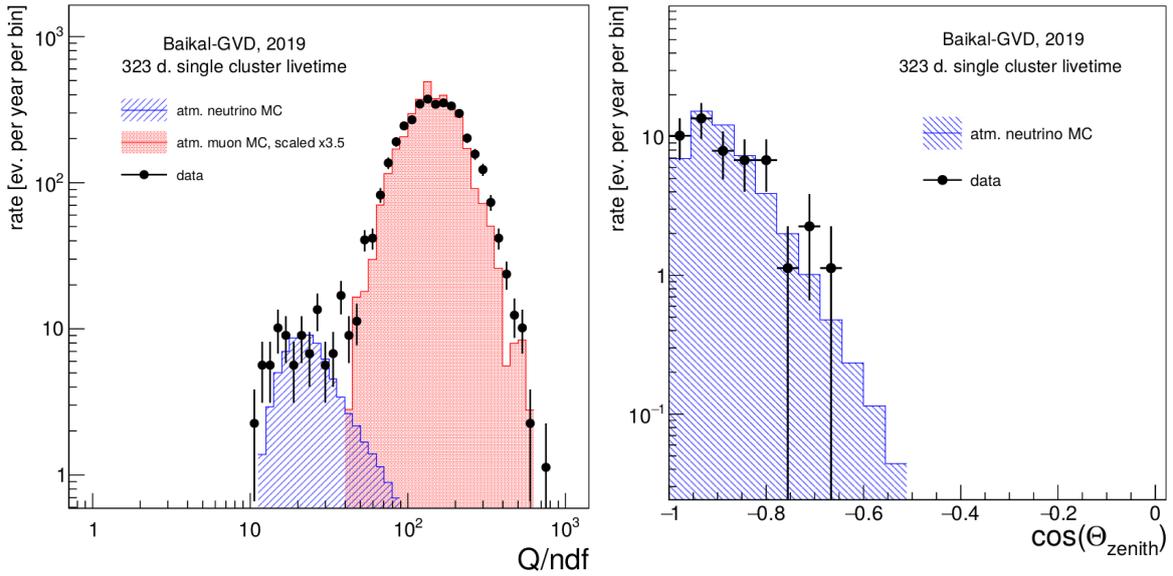

Figure 4.1.1 Left: Distribution of the fit quality parameter for tracks reconstructed as upward-going with cos θ < −0.5 in the single-cluster analysis. Experimental data are shown by black points with statistical error bars. The MC predictions for atmospheric muons and atmospheric neutrinos are shown in red and blue, respectively. Right: Reconstructed zenith angle distribution of the selected neutrino candidate events.

**4.2 High-energy cascades reconstruction**

The IceCube Neutrino Observatory has established the existence of a high-energy all-sky neutrino flux of astrophysical origin. This discovery was made using events interacting within a fiducial region of the detector surrounded by an active veto and with reconstructed energy above 60 TeV, commonly known as the high-energy starting event sample, or HESE [15]. The Baikal Collaboration has long-term experience with the NT200 array to search for diffuse neutrino fluxes via the cascade mode [16,17]. Search strategy for high-energy neutrinos with Baikal-GVD is based on the selection of cascade events generated by neutrino interactions in the sensitive volume of the array [18]. Here we discuss the preliminary results based on data accumulated in 2019-2020 (Zh.Dzhilkibaev, this conference, PoS(ICRC2021)1144).

To search for high-energy neutrinos of astrophysical origin the data collected by five clusters in 2019 and by seven clusters in 2020 have been used. A data sample accumulated by the array trigger, corresponds to 2915 one cluster live days. After applying procedures of cascade vertex and energy reconstruction and suppression of atmospheric muon background, seven events have been selected. Equal numbers of signal and background events are expected in experimental sample after applying final cuts. There is not statistically significant excess of total number of events was observed above expectation from atmospheric muon background. Cumulative energy distributions of experimental events (red dots), events expected from astrophysical flux with $E^{-2.46}$ spectrum and IceCube normalization (green histogram), events expected from atmospheric muons (brown histogram) and sum of expected astrophysical and atmospheric muon events (black histogram) are shown in Fig. 4.2.1. A higher statistics is required for observation of the diffuse astrophysical neutrino flux.

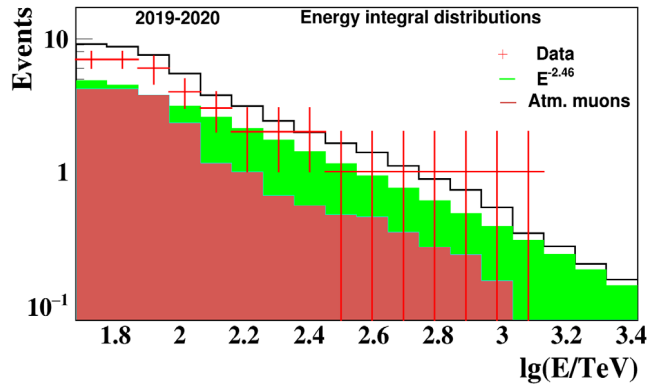

Figure 4.2.1 Cumulative energy distributions of experimental events (red dots), events expected from astrophysical flux with $E^{-2.46}$ spectrum and IceCube normalization (green histogram), events expected from atmospheric muons (brown histogram) and sum of expected astrophysical and atmospheric muon events.

Table 1: Parameters of the ten cascade events: reconstructed energy, zenith and azimuthal angles distances from cluster vertical axis, right ascension and declination.

|  | E,TeV | $\theta$, degree | $\varphi$, degree | $\rho$, m | R.A. | Dec. |
|---|---|---|---|---|---|---|
| GVD2018_1_354_N | 105 | 37 | 331 | 71 | 118.2 | 72.5 |
| GVD2018_1_383_N | 115 | 73 | 112 | 89 | 35.4 | 1.1 |
| GVD2018_1_656_N | 398 | 64 | 347 | 101 | 55.6 | 62.4 |
| GVD2019_1_114_N | 91 | 109 | 92 | 49 | 45.1 | -16.7 |
| GVD2019_2_112_N | 1200 | 61 | 329 | 96 | 217.7 | 57.6 |
| GVD2019_2_153_N | 129 | 50 | 321 | 52 | 33.7 | 61.4 |
| GVD2019_3_663_N | 83 | 50 | 276 | 73 | 163.6 | 34.2 |
| GVD2020_3_175_N | 110 | 71 | 185 | 84 | 295.3 | -18.9 |
| GVD2020_3_332_N | 74 | 92 | 9 | 19 | 223.0 | 35.4 |
| GVD2020_6_399_N | 246 | 57 | 49 | 80 | 131.9 | 50.2 |

Shown in Table 1 are parameters of the ten neutrino candidates – seven events selected from 2019-2020 data and three events with reconstructed energy E > 100 TeV selected from 2018 data and fulfilling all selection requirements: reconstructed energies, zenith and azimuthal angles as well as distances from cluster vertical axis and equatorial coordinates. Three, of ten astrophysical neutrino candidates are contained events. One event was reconstructed as a PeV scale cascade. Event GVD2019_1_114_N is reconstructed as upward going cascade with zenith angle θ = 109° and has very high probability to be neutrino induced cascade with signal-to-background ratio above 60%. Directional resolution of cascades strongly depends on cascade energy, vertex position and orientation with respect to array optical modules. In table 2 mismatch angles corresponding to 50%, 68%, 90% and 95% probability intervals are shown for the ten events.

The sky map of gamma-ray sources and $\Psi_{50\%}$ and $\Psi_{90\%}$ circles around the reconstructed positions of the ten neutrino candidates in galactic coordinates are shown in Fig. 4.2.2. Two events - GVD2018_1_656_N with 398 TeV reconstructed energy and GVD2019_2_153_N with 129 TeV energy were reconstructed close to Galactic Plane at a distance of 10° each other. In Fig. 4.2.3 left panel, zoom of sky map with two events are shown. There is gamma-ray active binary system LSI +61 303 which is covered by $\Psi_{95\%}$ circle of GVD2018_1_656_N and $\Psi_{50\%}$ circle of GVD2019_2_153_N. Binary system LSI +61 303 consist of massive Be star and unseen compact object which can be pulsar or black hole. Binary has orbital period $P_O$=26.496 days and superorbital period $P_{SO}$=1667 days. Emission from this system seen from radio to TeV gamma-rays. Last fact bring this system to rare class of TeV emitting binaries with only 4 objects known in Milky Way so far. Hadronic model for LSI +61 303 with prediction of neutrino flux was presented in [19].

Table 2: Mismatch angle intervals corresponding to 50%, 68%, 90% and 95% probabilities.

|  | $<\Psi_{50\%}$, degree | $<\Psi_{68\%}$, degree | $<\Psi_{90\%}$, degree | $<\Psi_{95\%}$, degree |
|---|---|---|---|---|
| GVD2018_1_354_N | 2.3 | 2.9 | 4.5 | 5.1 |
| GVD2018_1_383_N | 2.5 | 3.1 | 4.5 | 5.6 |
| GVD2018_1_656_N | 3.3 | 4.2 | 6.9 | 7.6 |
| GVD2019_1_114_N | 2.2 | 3.1 | 4.5 | 5.0 |
| GVD2019_2_112_N | 2.0 | 2.4 | 3.0 | 3.4 |
| GVD2019_2_153_N | 3.5 | 4.0 | 5.5 | 5.9 |
| GVD2019_3_663_N | 2.1 | 2.4 | 3.3 | 4.0 |
| GVD2020_3_175_N | 2.0 | 2.9 | 7.9 | 9.3 |
| GVD2020_3_332_N | 1.8 | 2.9 | 5.1 | 6.5 |
| GVD2020_6_399_N | 1.6 | 2.3 | 3.6 | 4.4 |

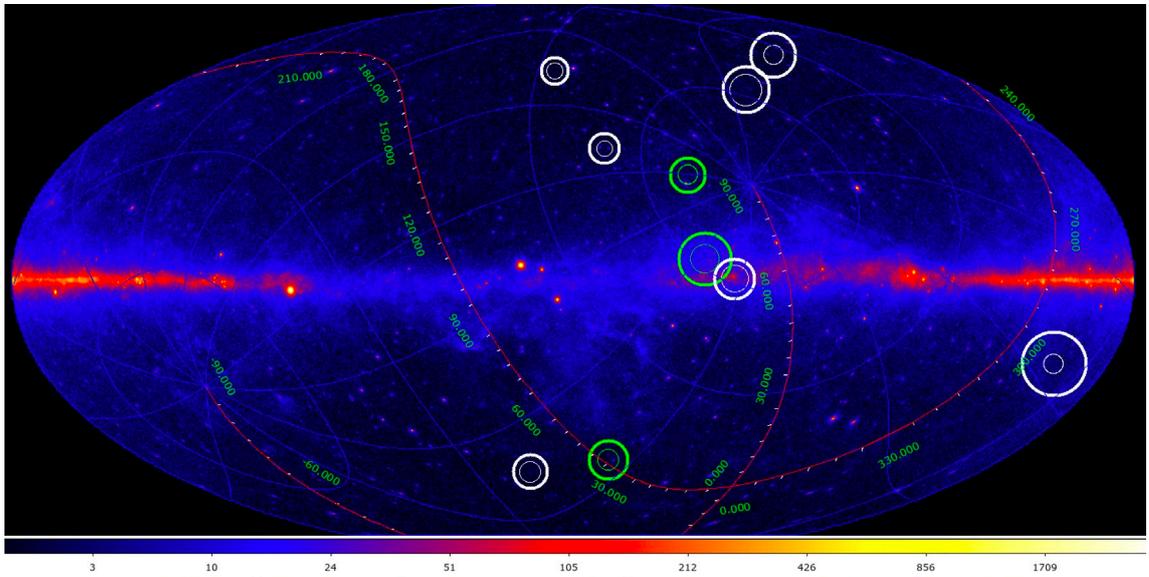

Figure 4.2.2 The sky map of gamma-ray sources with E > 1 GeV and $\Psi_{50\%}$ and $\Psi_{90\%}$ circles around the reconstructed positions of the ten neutrino candidates. Green circles relate to 2018 events.

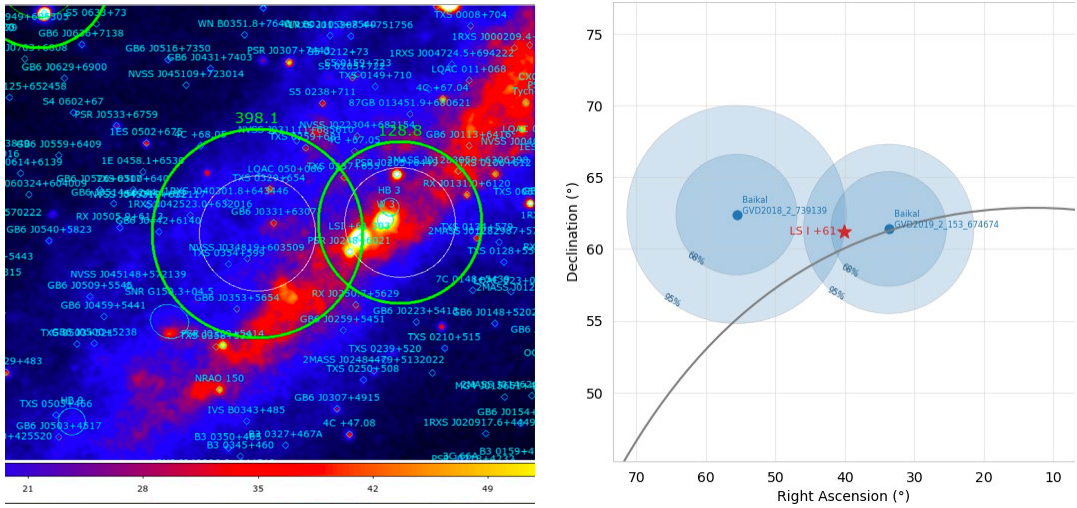

Figure 4.2.3 Left. The sky map of gamma-ray sources and $\Psi_{50\%}$ and $\Psi_{95\%}$ circles around the reconstructed positions of GVD2018_1_656_N and GVD2019_2_153_N. Right. The sky map with two neutrino events and gamma-ray active microquasar LSI +63 303.

Recent works have shown that high-energy neutrinos of TeV to PeV energies are produced in active galactic nuclei. Specifically, within several parsecs from the supermassive black hole of radio-bright blazars with jets pointing towards us [20,21]. We base the following analysis on those conclusions, and present our preliminary results regarding the blazar-neutrino connection. For each Baikal-GVD high-energy neutrino event, we first list all VLBI-bright blazars in the complete flux density-limited sample [22] falling within the 50% uncertainty region. These regions have radii of 2.5 or 4.5 degrees, and typically contain several blazars.

We follow [20] and attempt to select the most likely neutrino sources by analyzing the variability information in the form of radio light curves. Rich light curves are provided by monitoring campaigns at RATAN-600 [23] and OVRO [24] telescopes, and are available for a large subset of the VLBI sample. We visually identify blazars that undergo major radio flares close in time to the cascade event as potential neutrino sources. Two events have one nearby object experiencing a major flare. Light curves of these blazars are shown in Figures 4.2.4 and 4.2.5, together with sky maps surrounding corresponding GVD cascade events. The first is the J0301-1812 object. According to the multifrequency monitoring at RATAN-600, it was at the beginning of a major radio flare when a high-energy neutrino was detected from a direction 1.5 degree apart. The second example is J1938-1749, separated by 1.3 degrees from a cascade event. It was at the top of a flare, according to OVRO monitoring, and had the largest radio flux density observed over the whole monitoring duration.

These results are preliminary, the proper statistical significance of spatial and temporal associations is yet to be determined. Nevertheless, we find them promising and encourage following up with analysis of events and sources presented here, as well as future neutrino detections at Baikal-GVD.

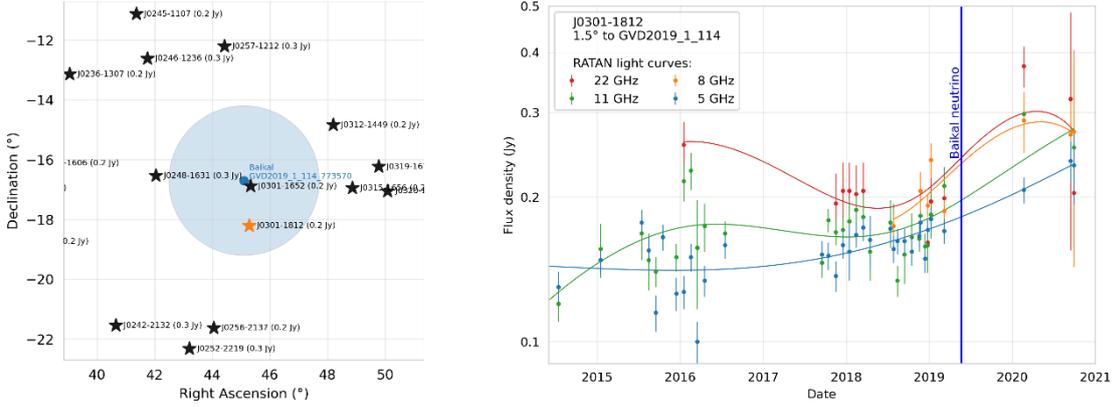

Figure 4.2.4 *Left*: Location of the GVD2019_1_114_N cascade event in the sky. The event itself is indicated by the blue dot in the center, with the light blue circle representing the 50% uncertainty region. VLBI-bright blazars from the complete sample are shown as stars with their names and radio flux densities to the side. The orange star corresponds to the blazar in the right panel. *Right*: Radio light curves from the RATAN-600 telescope for the J0301-1812 blazar, a potential neutrino source. This bright blazar undergoes a major flare, and the neutrino arrival coincides with the rise if this flare.

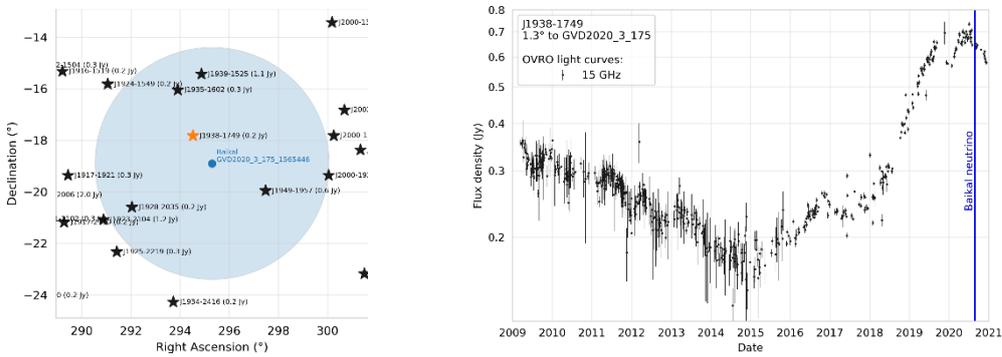

Figure 4.2.5 *Left*: Location of the GVD2020_3_175_N cascade event in the sky. See Fig. 4.2.4 caption for explanation of all symbols present. *Right*: The radio light curve from OVRO observations for the J1938-1749 blazar, a potential neutrino source. Neutrino arrived when this blazar was at its brightest over the whole observation period.

**4.3 Multi-messenger program at the Baikal-GVD telescope**

The Baikal-GVD participates in the international multi-messenger program on discovering the astrophysical sources of high energy fluxes of cosmic particles. Developing of online data processing, essential progress has been achieved in fall 2020, while at the beginning of 2021 the alert system of the Baikal-GVD detector has reached a quasi-online regime in fast reconstruction of runtime data collecting. Presently, the data processing mode where the processing is started when obtaining all files of the data accumulation session has been completely realized and is being used. The session processing duration and, accordingly, the delay in forming an alert event with given characteristics takes in average 3–5 hours (B.Shaybonov, this conference, PoS(ICRC2021)1040). An internal high-energy neutrino alert is formed among upgoing events as a result of the event reconstruction after the completion of a successive data transmission. The alerts are of two ranks of neutrino-like events: track of muon neutrino and very high energy cascade. In further development of the data collecting and processing we are going to form neutrino alarms to other communities.

The first HE alerts of the Baikal-GVD have been included in joint test of observing sample of the AGN of the 600 m RATAN radio telescope of the Special Astrophysical Observatory and the 40 m Telescope of the Owens Valley Radio Observatory (OVRO) (Zh.Dzhilkibaev, this conference, PoS(ICRC2021)1144).

It was our priority to follow up of the high-energy neutrino alerts of the ANTARES and the IceCube in searching for correlations in time and direction of the reconstructed GVD events with astrophysical alerts. The Baikal-GVD has received the ANTARES alerts of upward going muons in a real-time since December 2018. Following up of neutrino alert we look for events on each cluster in time windows ±500 sec, ±1 hour and ±1 day around alerts inside a half-open cone according to the GVD angular resolutions. In total, we have received 46 alerts and no prompt coincidence in time and direction has been found. However, for three alerts in 2019 we have found repeated cascades in time window ±1 day (O.Suvorova, this conference, PoS(ICRC2021)946). In Fig.4.3.1 (left) diurnal visibilities of these three alerts are shown. As have been found, in time window of ±1 day for coincidences of events in a cone of 5° the p-value to reject a hypothesis of background only events became about 3 sigma in each case. Further study of atmospheric muon background has been developed in a joint work of the ANTARES and GVD groups (S.A.Garre, PoS(ICRC2021)1121).

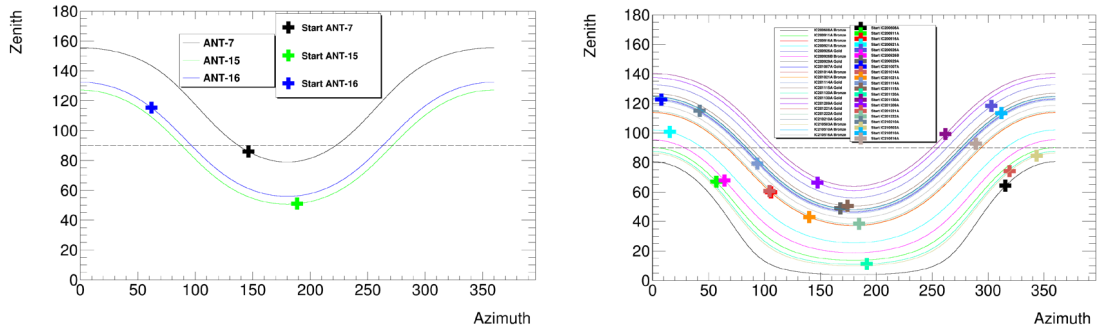

Figure 4.3.1 Diurnal trajectories of the fixed equatorial coordinates of neutrino alerts in Baikal-GVD horizontal coordinates. The crosses indicate alert locations. Left: the ANTARES events selected for further joint analysis due to GVD cascades follow-up. Right: The IceCube astrotracks in fall 2020. See text for details.

We began the follow-up analysis of the IceCube astrotracks of a hundred TeV energy in fast regime of data performance, by getting the GCN/AMON messages since August 2020 [25]. In Figure 4.3.1 (right) are shown locations of 21 astrotracks by crosses and their diurnal visibilities in the GVD horizon coordinates. It is seen that most part of astrotracks are observed at the GVD as down going events or near horizon events with zenith angles less than 120°. Thus, in prompt searches of event coincidences we used only cascade mode reconstruction. There were no events observed in the time window ±1 hour around each alert, while for few of them there were a cascades during larger time interval ±12 hours. In our recent work in [25] and in Figure 4.3.2 (left) it was shown that a level of the expected background in the time window ±12 hours varies between 0.29 and 0.45 events in cone of 5° around alerts. Therefore, there is no any statistically significant excess of the number of observed events above the expected background. The upper limits at 90% confidence level (c.l.) on number of events have been obtained for each IC alert according to Feldman and Cousins statistics [26]. Finally, constraints at 90% c.l. on the energy-dependent fluence of neutrinos of one type with a $E^{-2}$ spectrum under assumption of an equal fraction of neutrino flavors in the total fluence have been set for the IC alerts (see details in [25]). In Figure 4.3.2 (right) one can see reached level of the Baikal-GVD sensitivity to astrophysical neutrino sources of the Northern sky in the shown example for 9 IC astrotracks as dependence of the upper limits on neutrino fluence versus declination values.

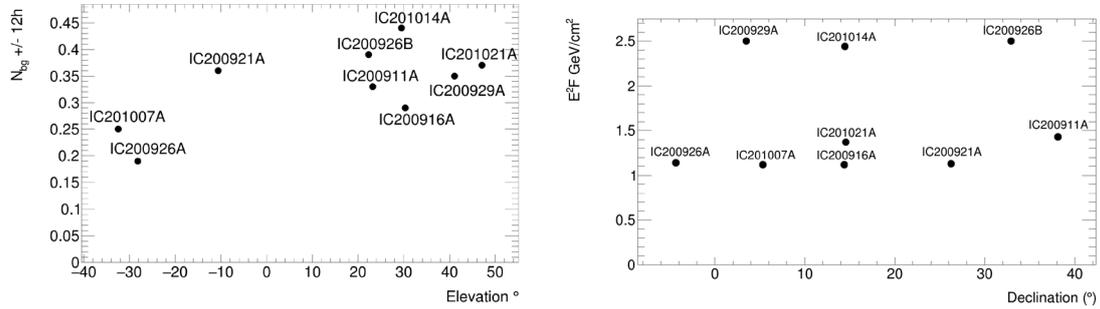

Figure 4.3.2: Left: Estimated number of background events for each IC alert in elevation coordinates. Right: The upper limits at 90% c.l. on neutrino fluence towards IC astrotracks of 2020 at given declinations.

In our off-line analysis of the Baikal-GVD data samples we have followed up two interesting cosmic phenomena: the optical transient AT2019dsg discovered by the Zwicky Transient Facility 2019-04-09 at 11:09:28.0 UTC with coordinates: α = 314.262° and δ = +14.204° and classified as a tidal disruption event [27] and the magnetar SGR 1935+2154 in the INTEGRAL discovery of its new activity in period of radio and gamma burst in April 27-28 of 2020 and CHINE/FBR observed radio burst 28 April 2020 14:34:33 UTC [28]. Among events reconstructed at five clusters towards the ATdsg2019 we have found 3 events at 3 different clusters with the delay time less than 12 hours between each other: MJD 58603.92444 (cluster 1), MJD 58603.82667 (cluster 3) and MJD 58603.95556 (cluster 4). Expected number of events has been obtained by scrambling of cascade data in right ascensions and a preliminary p-value was estimated as of 1.2% (2.26σ). Further statistical analysis is in progress now. The magnetar SGR 1935+2154 lies in the Galactic Plane (l = 57.25°, b = +0.82°) and could be associated with the supernova remnant G57.2+0.8 at distances less than 12.5 kpc. We follow up of this direction with data of the first 100 days after the INTEGRAL alert. In last level of cuts 2 cascades rests in 5 degrees around the source. Expected background was estimated as 0.35 from analysis of real data sample of a season 2016 and 2018. No excess in events was found, while the upper limits at 90% c.l. on number of events was obtained as 5.91 accordingly to statistics of [26]. Finally, the upper limits at 90% c.l. on the energy-dependent fluence of neutrinos of one type with a $E^{-2}$ spectrum under assumption of an equal fraction of neutrino flavors in the total fluence have been resulted in 2.0 GeV· cm$^{-2}$.

## 5. Conclusion

The ultimate goal of the Baikal-GVD project is the construction of a km$^3$-scale neutrino telescope with the implementation of about ten thousand photo-detectors. The first cluster in its baseline configuration was deployed in 2016. In total, eight clusters with 2304 OMs arranged at 64 strings are data taking since April 2021. Baikal-GVD in 2021 configuration is the largest neutrino telescope in the North at present. The modular structure of the Baikal-GVD design allows for studies of neutrinos of different origin at early stages of the construction. The analysis of a combined dataset of single-cluster events collected from the first five operational clusters of Baikal-GVD over a two monthes period in spring 2019, yields 44 upward going neutrino events from atmospheric neutrino flux. The observed flux and angular distribution of the neutrino events is in good agreement with Monte Carlo predictions. The analysis of data collected in 2018 and 2020 allows the selection of ten promising high-energy cascade events - candidates for events from astrophysical neutrinos. Multi-messenger studies started in 2017 with searching for neutrino signal associated with GW170817. Fast follow-up searches for coincidences of Baikal-GVD events with ANTARES/TAToO and IceCube high energy neutrino are performing sience 2019. The alarm system in real-time monitoring of the celestial sphere was launched at the beginning of 2021, that allows to form the GVD alerts of two ranks like "muon neutrino" and "VHE cascade". The commissioning of the first stage of the Baikal neutrino telescope GVD-I with an effective volume 0.8 km$^3$ is envisaged for 2024-2025.

*We acknowledge the support by the Ministry of Science and Higher Education of Russian Federation under the contract FZZE-2020-0017. The work was supported by RFBR grant 128 20-02-00400. The CTU group acknowledges the support by European Regional Development Fund-Project No. CZ.02.1.01/0.0/0.0/16_019/0000766.*

## References

[1] The IceCube, Fermi-LAT, MAGIceCube, AGILE, ASAS-SN, HAWC, H.E.S.S, INTEGRAL, Kanata, Kiso, Kapteyn, Liverpool telescope, Subaru, Swift/NuSTAR, VERITAS, and VLA/17B-403 teams, Science **361**, 1378 (2018).

[2] Aartsen M.G. et al. (IceCube Coll.) Science 361, 147-151 (2018).

[3] A.V. Avrorin et al. (Baikal Collaboration), *The Gigaton volume detector in Lake Baikal*, Nucl. Instr. and Meth. in Phys. Res. A, **639** (2011) 30.

[4] M. G. Aartsen et al., *The IceCube Neutrino Observatory: instrumentation and online systems*, JINST 12 (2017) P03012.

[5] S. Adrian-Martinez et al., *KM3NeT 2.0 - Letter of intent for ARCA and ORCA*, J. Phys. G 43 (2016) 084001.

[6] A.D. Avrorin et al., *Spatial positioning of underwater components for Baikal-GVD*, EPJ Web of Conf. 207 (2019) 07004.

[7] A.D. Avrorin et al., *The Baikal-GVD detector calibration*, PoS-ICRC2019-0878, arXiv:1908.05458.

[8] A.D. Avrorin et al., *Time calibration of the neutrino telescope Baikal-GVD*, EPJ Web of Conf. 207 (2019) 07003.

[9] A.D. Avrorin et al., *Search for cascade events with Baikal- GVD*, PoS-ICRC2019-0873, [arXiv:1908.05430]

[10] V. Balkanov et al., *In situ measurements of optical parameters in Lake Baikal with the help of a Neutrino telescope*, Applied Optics Vol. 38, Issue 33, pp. 6818-6825 (1999) Appl. Opt. 38 (1999) 6818.

[11] A.D. Avrorin et al., *Asp-15: A stationary device for the measurement of the optical water properties at the NT200 neutrino telescope site*, Nucl. Instrum. Meth. A 693 (2012) 186.

[12] V.M. Aynutdinov et al. (Baikal Collaboration), *The data acquisition system for Baikal-GVD*, EPJ Web of Conferences, **116** (2016) 5004.

[13] J. Serrano et al, *The White Rabbit Project*, Proceedings of ICALEPS 2009, ICALEPCS TUC004 (2009)**.**

[14] V.A. Allakhverdyan et al., *Measuring muon tracks in Baikal-GVD using a fast reconstruction algorithm*, submitted to Eur. Phys. J. C, [arXiv:2106.06288].

[15] I. Taboada, Talk at XXVIII International Conference on Neutrino Physics and Astrophysics, 4-9 June 2018, Heidelberg, Germany, DOI: 10.5281/zenodo.1286918,
URL: https://doi.org/10.5281/zenodo.1286918, [arXiv:2011.03545].

[16] V. Aynutdinov et al., Astropart.Phys. **25**, 140 (2006).

[17] A.V Avrorin et al., Astronomy Letters, **35** 651 (2009).

[18] A.D. Avrorin et al., PoS (ICRC2017)962, (2017).

[19] A.Neronov and M.Ribordy, Phys. Rev. D 79, 043013 (2009)

[20] A. Plavin et al., ApJ 894, 101 (2020).

[21] A.Plavin et al., ApJ 908, 157 (2021).

[22] The Radio Fundamental Catalog, http://astrogeo.org/rfc/

[23] Y.Y. Kovalev et al., PASA 19, 83 (2002).

[24] J. Richards et al., Astrophys. J. Suppl. 194, 29 (2011).

[25] A.D. Avrorin et al., Astronomy Letters, 47, 94-104 (2021).

[26] G. Feldman and R. Cousins, Phys. Rev. D 57, 3873 (1998).

[27] https://www.wis-tns.org/object/2019dsg

[28] S.Mereghetti et al., https://iopscience.iop.org/article/10.3847/2041-8213/aba2cf, (2020).

**Full Authors List: BAIKAL-GVD Collaboration**

V.A. Allakhverdyan[1], A.D. Avrorin[2], A.V. Avrorin[2], V.M. Aynutdinov[2], R. Bannasch[3],

Z. Bardačová[4], I.A. Belolaptikov[1], I.V. Borina[1], V.B. Brudanin[1], N.M. Budnev[5],

V.Y. Dik[1], G.V. Domogatsky[2], A.A. Doroshenko[2], R. Dvornický[1,4], A.N. Dyachok[5],

Zh.-A.M. Dzhilkibaev[2], E. Eckerová[4], T.V. Elzhov[1], L. Fajt[6], S.V. Fialkovski[7], A.R. Gafarov[5],

K.V. Golubkov[2], N.S. Gorshkov[1], T.I. Gress[5], M.S. Katulin[1], K.G. Kebkal[3], O.G. Kebkal[3],

E.V. Khramov[1], M.M. Kolbin[1], K.V. Konischev[1], K.A. Kopański[8], A.V. Korobchenko[1],

A.P. Koshechkin[2], V.A. Kozhin[9], M.V. Kruglov[1], M.K. Kryukov[2], V.F. Kulepov[7], Pa. Malecki[8],

Y.M. Malyshkin[1], M.B. Milenin[2], R.R. Mirgazov[5], D.V. Naumov[1], V. Nazari[1], W. Noga[8],

D.P. Petukhov[2], E.N. Pliskovsky[1], M.I. Rozanov[10], V.D. Rushay[1], E.V. Ryabov[5],

G.B. Safronov[2], B.A. Shaybonov[1], M.D. Shelepov[2], F. Šimkovic[1,4,6], A.E. Sirenko[1],

A.V. Skurikhin[9], A.G. Solovjev[1], M.N. Sorokovikov[1], I. Štekl[6], A.P. Stromakov[2],

E.O. Sushenok[1], O.V. Suvorova[2], V.A. Tabolenko[5], B.A. Tarashansky[5], Y.V. Yablokova[1],

S.A. Yakovlev[3], D.N. Zaborov[2]

[1] Joint Institute for Nuclear Research, Dubna, Russia, 141980

[2] Institute for Nuclear Research, Russian Academy of Sciences, Moscow, Russia, 117312

[3] EvoLogics GmbH, Berlin, Germany, 13355

[4] Comenius University, Bratislava, Slovakia, 81499

[5] Irkutsk State University, Irkutsk, Russia, 664003

[6] Czech Technical University in Prague, Prague, Czech Republic, 16000

[7] Nizhny Novgorod State Technical University, Nizhny Novgorod, Russia, 603950

[8] Institute of Nuclear Physics of Polish Academy of Sciences (IFJ PAN), Krakow, Poland, 60179

[9] Skobeltsyn Institute of Nuclear Physics MSU, Moscow, Russia, 119991

[10] St. Petersburg State Marine Technical University, St. Petersburg, Russia, 190008